\documentstyle[12pt,jeep]{article}
\numberbysection
\begin{document}
\title{\bf Excitation Thresholds for Nonlinear Localized Modes on Lattices}
\author{
  M.I. Weinstein \thanks{Department of Mathematics,
  University of Michigan, Ann Arbor, MI and Mathematical Sciences Research, 
 Bell Laboratories - Lucent Technologies, Murray Hill, NJ
  }}
   \baselineskip=18pt
   \maketitle
\def\M{\Gamma^{-\alpha}}
\def\un{\underline}
\def\nn{\nonumber}
\newcommand{\no}{\nonumber}
\newcommand{\be}{\begin{equation}}
\newcommand{\ee}{\end{equation}}
\newcommand{\ba}{\begin{eqnarray}}
\newcommand{\ea}{\end{eqnarray}}
\newcommand{\ve}{\varepsilon}
\newcommand{\nit}\noindent
\newcommand{\D}\partial
\def\Pc{{\bf P_c}}
\def\a{\alpha}
\def\b{\beta}
\newcommand{\lan}\langle
\newcommand{\ran}\rangle
\newcommand{\wpl}{w_+}
\newcommand{\wmi}{w_-}
\newcommand{\vpsi}{\vec\psi}
\newcommand{\vg}{\vec g}
\newcommand{\la}{\lambda}
\newcommand{\ra}{\rightarrow}
\newcommand{\Z}{{\rm Z\kern-.35em Z}}
\newcommand{\bP}{{\rm I\kern-.15em P}}
\newcommand{\Q}{\kern.3em\rule{.07em}{.65em}\kern-.3em{\rm Q}}
\newcommand{\R}{{\rm I\kern-.15em R}}
\newcommand{\h}{{\rm I\kern-.15em H}}
\newcommand{\C}{\kern.3em\rule{.07em}{.55em}\kern-.3em{\rm C}}
\newcommand{\T}{{\rm T\kern-.35em T}}
\newtheorem{theo}{Theorem}[section]
\newtheorem{defin}{Definition}[section]
\newtheorem{prop}{Proposition}[section]
\newtheorem{lem}{Lemma}[section]
\newtheorem{cor}{Corollary}[section]
\newtheorem{rmk}{Remark}[section]
\begin{abstract}
\nit
Breathers are spatially localized and time periodic solutions of 
extended Hamiltonian dynamical systems.
In this paper we study  {\it excitation thresholds} for (nonlinearly 
dynamically
stable) ground state  
breather or standing wave solutions for  
networks of coupled nonlinear oscillators and wave equations of nonlinear
Schr\"odinger (NLS) type.
Excitation thresholds are rigorously characterized by variational methods. The  
excitation threshold is related to the optimal (best) constant in a class of discrete 
interpolation inequalities related to the Hamiltonian energy. 
We establish a precise connection among $d$, the dimensionality of the lattice,
$2\sigma+1$, the degree of the nonlinearity and  the existence of an excitation
threshold
for discrete nonlinear Schr\"odinger systems (DNLS).
 We prove that if $\sigma\ge{2\over d}$, then ground state standing waves exist
  if and only if
 the {\it total power} is larger than some strictly positive threshold,
 $\nu_{thresh}(\sigma, d)$. This proves a conjecture of Flach, Kaldko
 \& MacKay \cite{kn:FKM} in the context of DNLS.
  We also discuss upper and lower bounds for excitation thresholds
 for ground states of coupled systems of NLS  equations, which arise in the modeling  of 
 pulse propagation in coupled arrays of optical fibers.
\end{abstract}
\thispagestyle{empty}
%
\section{Introduction}

  This article concerns threshold behavior of certain time-reversible, energy preserving
   nonlinear dynamical systems. 
  Consider an infinite dimensional 
  Hamiltonian system (wave equation or network of discrete oscillators)
  defined on an infinite spatial domain. If the system is translation invariant
  ( {\it e.g.}, not having any localized potential wells), 
  one expects that "small amplitude" or "low energy" 
   solutions will disperse to zero; see, for example, \cite{kn:Strauss}. If the
  system is nonlinear and having an {\it attractive} nonlinear
  potential, one can expect that sufficiently large "amplitude" initial data will lead to an
  evolution consisting of a non-decaying "bound state" component 
  plus a dispersive component ({\it radiation}), which tends (weakly) to zero with
  increasing time.
  In this latter scenario, we think of permanent non-decaying structures as
  having been excited by the initial condition; a deep enough self-consistent
  potential well has been initialized in which one can sustain a permanent
  structure.
   Since the systems we are discussing are infinite
  dimensional, the sense in which one measures amplitude is crucial. In systems
  of physical interest, there is often a natural 
   measure of a solution's size. Roughly speaking, 
   if there is a critical size,  
   $\nu_{thresh}>0$, such   
  that there are permanent (non-decaying in time) states of size $\nu$ 
   if and only if
  $\nu>\nu_{thresh}$, then we refer to
  $\nu_{thresh}$ as an {\it excitation threshold}.
 In this paper, we  investigate the existence and nonexistence of  excitation thresholds for a class of time-periodic and spatially
 localized standing wave  
 states for two classes of dynamical systems.  In certain models, these states
 have been called "breathers". See section 3 for a precise definition of
 and discussion concerning excitation
 thresholds. 
The dynamical systems we consider are: (1) the discrete nonlinear Schr\"odinger
equation (\ref{eq:dnls}) and 
 (2) a system of coupled nonlinear Schr\"odinger equations (\ref{eq:cnls}); see also
 (\ref{eq:cnls1}).

 Mathematical models which support
discrete breathers are of interest in the study of vibrations in, for
example, localized crystals and biological molecules 
 \cite{kn:Eilbecketal}, \cite{kn:FW}. Recently, experimental observations of 
  such discrete nonlinear 
localized modes have been made in 
  coupled systems of optical waveguides \cite{kn:experiment}. 
With a view toward study of
such structures in experiment it is of interest to understand under what
circumstances a discrete breather is excited. 

In \cite{kn:MW} a formal variational argument is given suggesting the
existence of such energy thresholds for the one-dimensional
 discrete nonlinear Schr\"odinger
(DNLS) equation (also known as the discrete self-trapping equation
\cite{kn:Eilbecketal}).
For the related system of nearest neighbor coupled nonlinear Schr\"odinger
equations, (\ref{eq:cnls}), such thresholds were rigorously demonstrated to exist
\cite{kn:WY}.

 In the recent paper of Flach, Kladko \& MacKay \cite{kn:FKM}, 
  heuristic scaling arguments and
 numerical studies are presented which suggest that for a large class of
  Hamiltonian dynamical systems defined on one, two and three
 dimensional lattices, there is a lower bound on the energy of a breather if
 the lattice dimension is greater than or equal to a certain critical
 value. 
 %
 %

 Theorem 3.1 resolves this conjecture for 
  ground state breathers of the  
  $d$- dimensional 
 discrete 
  nonlinear Schr\"odinger equation (DNLS):
	 \be
	  i\D_t\psi_l\ =\ -\kappa\ \left(\ \delta^2\ \vec\psi\ \right)_l\
	-\ |\psi_l |^{2\sigma}\ \psi_l,\ \ \kappa>0
	\label{eq:dnls}
	\ee
Here,
 $ {\vpsi}\ =\ \{\psi_l(t)\},\ \ l\in \Z^d,\ \ t\in\R$, $\delta^2$ denotes the
 $d$-dimensional discrete Laplacian on $\Z^d$
  given by:
  \be
\left(\ \delta^2\vec\psi\ \right)_l\ =\ \sum_{m\in N_l}\psi_m\ -\ 2d\
\psi_l,
\label{eq:discreteLaplacian}
\ee
where $N_l$ denotes the set of $2d$ nearest neighbors of the point in
$\Z^d$ with label $l$. The parameter $\kappa$ can be interpreted as a
discretization parameter, $\kappa\ \sim\ h^{-2}$, where $h$ is the lattice
spacing and $\psi_l\ =\ \psi(hl),\ l\in\Z^d$. 
The parameter $\sigma>0$ is a measure of the degree  
of nonlinearity. 

Theorem 3.1 states that there is a ground
state $l^2$- excitation threshold if and only  if $\sigma\ge{2\over d}$.
For $\sigma<{2\over d}$ breathers of arbitrarily small $l^2$ norm exist. 
See \cite{kn:Flach1}, \cite{kn:Flach2}, \cite{kn:FKM} for a study of the bifurcation of small amplitude states from the edge of the plane wave spectrum.
In contrast, the continuum limit nonlinear Schr\"odinger
equation, (\ref{eq:nls}), has an $L^2$ threshold only in the case
 of critical nonlinearity, $\sigma = {2\over d}$. This is a manifestation
 of the role of discreteness, which
 breaks the dilation invariance of the continuum case; see the discussion and
 analysis of sections 3 and 4. 
  Theorem 2.2 states that ground states 
are nonlinearly
dynamically stable in an orbital sense; see also \cite{kn:LST}. 

 In section 5 we consider
the limiting behavior of ground states of {\it total power}: 
 $ \|\vec \psi\|_{l^2}^2\ =\ \nu$, as $\nu$ tends to infinity. Such ground states
 are found to have large amplitude. As $\nu$ is increased they are
  increasingly concentrated about one lattice site. 
   A phenomenon of this type has been
  observed for the systems (\ref{eq:cnls}), (\ref{eq:cnls1}),  
   and analytically studied in \cite{kn:WY}, \cite{kn:yearythesis}.
 The relation of this result to the numerical work of
 \cite{kn:BRC} and to the work on the {\it anti-integrable limit}
 \cite{kn:MA}, \cite{kn:Aubry} is also discussed.

Studies of discrete breathers originated in the context of classical
nonlinear wave equations. An example is the one-dimensional Klein-Gordon
equation:
\be
\D_t^2\ u_n\ =\ D\ \left(\ u_{n+1}\ -\ 2\ u_n\ +\ u_{n-1}\ \right)\
-\Omega_0^2\ u_n\ +\ u_n^3.\label{eq:nlkg}
\ee
The techniques of this paper do not directly apply to give rigorous
thresholds for discrete 
 nonlinear Klein Gordon equation localized states. However,
our results concerning  DNLS are related, through a 
 multiple scale approximation, appropriate to the limit of 
large lattice spacing, $h$. 
 Specifically, let $h\ =\ \kappa^{-{1\over2}}\ \varepsilon^{-1}\ $,
  and therefore  
  $D\ =\ \varepsilon^2\ \kappa$.  
Then, seeking a solution of the form:
\be
u_n\ =\ \varepsilon\ \Psi_n+\ \varepsilon^2\ \Phi_n\ +\ \varepsilon^3\
\chi_n\ +\ ...
\label{eq:un-ansatz}\ee
we find an approximate solution which is valid for times, $t$,  of order
$\varepsilon^{-2}$ with 
\be
\Psi_n\ =\ \Psi_n(t,T)\ =\ e^{-i\Omega_o t}\ \psi_n(T)\ + 
 e^{i\Omega_o t}
\ \overline{\psi_n}(T),\ T={1\over2} \varepsilon^2 t
\nn\ee
where $\psi_n(T)$ satisfies the discrete nonlinear Schr\"odinger equation
(\ref{eq:dnls}). In particular, this yields using the results of this
paper on DNLS approximate solutions of the form:
\be
u^\varepsilon_n(t;\omega)\ 
 =\ 2\ \varepsilon\ \cos\left(\ [\Omega_0 + \varepsilon^2\omega ] t +
\gamma\ \right) g_n\ +\ {\cal O}(\varepsilon^2),
\nn\ee
where $\omega<0$ and $\vec g\ =\ \vec g_\omega\ =\
  \{\ g_n\ \}_{n\in\Z}\
\in\ l^2(\Z)$.

Finally, in section  6 we  discuss and extend results on 
 excitation thresholds for ground
states of a class of 
 coupled system of nonlinear
Schr\"odinger equations (CNLS), which arises in the modeling of pulse propagation
through a coupled network of optical fibers \cite{kn:Acevesetal}, \cite{kn:ALDRT},
 \cite{kn:LSTM}, \cite{kn:BA}
\cite{kn:WY}, \cite{kn:yearythesis}:
\ba
i\D_t\ \psi_l\ +   \ \D_x^2\psi_l\ +\
 \ \kappa\ \left(\ \delta^2\ \vec\psi\ \right)_l\ 
  +\ 2\ |\psi_l|^2\ \psi_l\ &=&\ 0,
\label{eq:cnls}\\
\vec\psi\ =\ \{\psi_l(t,x)\}_{l\in\Z^d},\ \ (t,x)&\in&\R^2\nn
\ea

The cases of
physical interest are $d=1,2$.
Here, $\psi_l$ denotes the slowly varying envelope of the highly oscillatory
electric field in the fiber with position $l$ in the lattice. 
 We consider the case where $l$ varies over
$\Z^d$, with $\sum_l\ \|\psi_l\|^2_{L^2(\R)}<\infty$. For the case $\sigma=d=1$, we
obtain numerical upper and lower bounds (\ref{eq:bounds}) for the excitation
thresholds $\nu_c$.\footnote{An error in these bounds due to faulty algebra
appeared in \cite{kn:WY} and is corrected here.} 
 Other boundary conditions are discussed in \cite{kn:WY}, \cite{kn:yearythesis}. In
particular, a result of the analysis is that there are no excitation thresholds
in the case when the system is periodic in the discrete variable, $l$; 
 ground states of arbitrary positive total power $\nu = \sum_l\
\| \psi_l \|_2^2$ exist.

In this paper we use observations about the scaling
structure of variational problems together with 
compactness methods in the calculus of
variations; see {\it e.g.} \cite{kn:PLL}, \cite{kn:BL}. Thresholds for the excitation of
breathers or nonlinear bound states are characterized in terms of the optimal
(best) constant of discrete interpolation inequalities for elements of
$l^2(\Z^d)$ in the case of (\ref{eq:dnls}) and for elements of 
 $l^2(\Z^d;H^1(\R))$ in
the case of (\ref{eq:cnls}). This is related to the approach taken in
\cite{kn:nls83}, \cite{kn:survey}
 on a sharp criterion on initial conditions for global existence (no
 finite time 
 blow-up) of
solutions to the continuum nonlinear Schr\"odinger equation
 on $\R^d$, (\ref{eq:nls}), with critical power
nonlinearity. Results on excitation thresholds, stability and other issues 
 for the semi-discrete class of nonlinear Schr\"odinger
equations were obtained by B. Yeary and the author \cite{kn:WY}, \cite{kn:yearythesis}. 
 This article is  a detailed account with extensions 
  of the work on excitation thresholds.

 {\bf Acknowledgements:}
 The author wishes to thank Peter D. Miller for very
 helpful comments. 
  This work was supported in part by a grant from the U.S. National
 Science Foundation.

\section{ DNLS and a variational characterization of its ground state}

By standard methods, one can check that for any 
$\vpsi(t=0)\in l^2(\Z^d)$, there is a unique global solution $\vpsi\in
C^1(\R;l^2(\Z^d))$ of DNLS, (\ref{eq:dnls}), 
 and for which the following two quantities are
independent of time:
\ba
{\cal H}_D\left[\ \vpsi\ \right]\ &=&\ 
 -\kappa\left(\ \delta^2\vpsi,\vpsi\ \right)
 \  -\ {1\over{\sigma+1}} \sum_{l\in \Z^d}
 |\psi_l|^{2\sigma +2},
 \label{eq:Hamiltonian}\\
 {\cal N}_D\left[\ \vpsi\ \right]\ &=&\ \sum_{l\in \Z^d}  |\psi_l|^2.
\label{eq:L2norm}
\ea
The subscript, "$D$", is used to indicate a quantity associated with the discrete
equation (\ref{eq:dnls}).
${\cal H}_D$ is a Hamiltonian for (\ref{eq:dnls}), which can be written
as:
\be
i\D_t\vec\psi\ =\ {\delta{\cal H}_D\over{\delta\vec\psi^*}}.
\nn\ee
In various applications the invariant ${\cal N}$ has the interpretation of {\it total
power} or of {\it particle number}.
The term $-\left(\delta^2\vpsi,\vpsi\right)$ may be written out explicitly
as:
\be
\left(\ -\delta^2 \vpsi,\vpsi\ \right)\ =\ \sum_{r=1}^d\ \sum_{l\in\Z^d}
 |\psi_l -\psi_{\tau_r l}|^2,
 \nn\ee
 where $\tau_r$ denotes translation by one lattice unit in the $r^{th}$
 coordinate direction.

 Of particular interest are  spatially localized and time-periodic 
 solutions. We seek them in the form:
 \ba
 \psi_l(t)\ &=&\ e^{-i\omega t}\ g_l,\ \ l\in \Z^d,\ t\in\R,\nn\\
 \psi_l(t)\ &\in&\ l^2{(\Z^d)}.
 \label{eq:boundstate}
 \ea
 where $\omega$ is real.
  A solution of this type is frequently called a  nonlinear 
  {\it bound state, standing wave}  or {\it stationary state}. The term 
 {\it discrete breather} is also used but is sometimes reserved for
 a localized state whose modulus oscillates.

 Substitution of (\ref{eq:boundstate}) into (\ref{eq:dnls}) yields the
 system of algebraic equations plus "boundary condition at infinity":
 \ba
 \omega\ g_l\ &=&\ -\kappa\ \left(\delta^2\ g\right)_l\ -\ |g_l|^{2\sigma}\
 g_l.
 \label{eq:gequation}\\
 \vg\ &=&\ \{ g_l\}_{l\in\Z^d}\ \in\ l^2(\Z^d).
 \label{eq:ginl2}\ea

We construct a ground state by variational methods. To
motivate our approach, we consider the quantum mechanical problem:
\be
H\ \Psi\ =\ E\ \psi,\nn
\ee
where $H= -\Delta + V(x)$ for a bound state, $\Psi\in L^2$ with
$\|\Psi\|_2=1$. We assume  $V(x)$ is a sufficiently smooth and rapidly
decaying "potential well". Consider the constrained minimization problem:
\be
{\cal I}\ =\ \inf\{\ (Hf,f)\ :\ \|f\|_2=1\ \}.
\label{eq:Idef}
\ee
If ${\cal I}<0$, then $E_g\equiv {\cal I}$ is the ground state (lowest) 
 eigenvalue and there exists a ground state eigenstate $\Psi_g(x)$ such
 that 
 \be
 H\Psi_g\ =\ E_g\ \Psi_g,\ \|\Psi_g\|_2=1.
 \nn\ee
 The time-periodic {\it breather} or standing wave, $\Psi_g(x)e^{-iE_gt}$, is a
 dynamically stable solution of the time-dependent Schr\"odinger equation
 \be
 i\D_t\ \Psi\ =\ H\ \Psi.
 \label{eq:schroedinger}
 \ee
 We shall characterize the ground state of (\ref{eq:dnls}) using a 
  nonlinear analogue of (\ref{eq:Idef}).
\bigskip

  \nit {\bf Definition:} Let 
  \be {\cal I}_\nu\ =\ 
  \inf\{ {\cal H}_D[\ \vec f\ ]\ :\ {\cal N}_D[\ \vec f\ ]\ =\ \nu \}.
  \label{eq:Inudef}
  \ee
  A minimizer of the variational problem (\ref{eq:Inudef}) is called a
  {\it ground state}.
\medskip

  Clearly, ${\cal I}_\nu$ is bounded below: For,
   \be
   {\cal H}_D[\ \vec f\ ]\ \ge -{1\over\sigma +1}\sum_{l}|f_l|^{2\sigma+2}\ge
	-{1\over\sigma +1}\|\ \vec f\ \|_\infty^{2\sigma}\ \|\ \vec f\ \|_2^2\ge 
	-{1\over\sigma +1}\ \nu^{\sigma+1}.\nn\ee

\begin{theo}

\nit (a)\ If $-\infty\ <\ {\cal I}_\nu\ <\ 0$, then the minimum in
 (\ref{eq:Inudef}) is attained. 

\nit (b)\ Every minimizing sequence associated with the variational problem
(\ref{eq:Inudef}) is precompact modulo phase translations, {\it i.e.} for any minimizing
sequence $\{\ \vec g^{(k)}\ \}$, there is a subsequence $\{\ \vec g^{(n_k)}\ \}$ and a
sequence $\{\ \gamma_{n_k}\ \}$, and translations, $\tau(l^k)$ 
(where $\tau(l^k) \vec
 g^{(k)}\ =\ \{\ g^{(n_k)}_{j+l^k}\}_{j\in\Z^d}$),  such that $\tau(l_k)\vec g^{(n_k)}\
e^{i\gamma_{n_k}}$ converges in $l^2(\Z^d)$ to a minimizer.

\nit (c)\ If $\vg\ =\ \{g_l\}_{l\in\Z^d}$ is a minimizer for the variational
problem (\ref{eq:Inudef}), then there
exists $\omega=\omega(\nu)<0$ such that the Euler-Lagrange equation:
\be
\omega(\nu)g_l\ =\ -\kappa\ \left( \delta^2\ g\right)_l\ -\ |g_l|^{2\sigma}\ g_l,
\ l\in\Z^d
\label{eq:gEL}\ee
holds, together with the $L^2$ constraint:
\be
{\cal N}_D\left[\ \vg\ \right]\ =\ \sum_l |g_l|^2\ =\ \nu.
\nn\ee
\end{theo}

This theorem can be proved by a standard application of concentration
compactness ideas in the discrete context
\cite{kn:PLL}; see \cite{kn:yearythesis}. An outline of the proof is presented in  
 appendix A.
 \bigskip

\centerline{\it Dynamical stability}

 Before stating a precise result, we first introduce
some terminology and notation.

\nit{\bf Definitions:}

\nit (1) Let ${\cal G}_\nu$ denote the set of all solutions of the minimization from
(\ref{eq:Inudef}), {\it i.e.} the set of ground states with ${\cal N}\ =\ \nu$.

\nit (2) Given a particular ground state $\vec g$, we define its {\it orbit} to be the set:
\be
{\cal O}(\ \vec g\ )\ =\ \{ e^{i\gamma}\ \vec g\ :\ \gamma\in [0,2\pi)\}
\label{eq:orbit}
\ee

\nit (3) The {\it distance} $\rho\left(\ \vec\psi\ ,\  {\cal G}_\nu \ \right)$ 
 from 
 $\psi\in l^2$ to the set of ground states, ${\cal G}_\nu$
is given by:
\be
\rho\left(\ \vec\psi\ ,\ {\cal G}_\nu \right)\ 
 \equiv\ \inf_{\vec g\in {\cal G}_\nu } 
 \|\ \vec\psi\ -\ \vec g\ \|_{l^2(\Z^d)}
\label{eq:rhodef}
\ee
\medskip

\nit{\bf Remark:} We conjecture that the ground state with ${\cal N}=\nu$
is essentially unique, {\it i.e.} if  $\vec g$ is any ground state with 
${\cal N}[\ \vec g\ ]\ =\nu$, then ${\cal G}_\nu = {\cal O}(\vec g)$.

A consequence of part (b) of Theorem 2.1 is the following \cite{kn:CL}:
\begin{theo}
Ground states of (\ref{eq:dnls})
 are  {\it orbitally Lyapunov stable} in the sense that: given any
$\varepsilon>0$, there is a $\delta>0$ such that if the initial data $\vec \psi(t=0) =
\vec\psi_0$ satisfies 
\be
\rho\left(\ \vec\psi_0\ ,\ {\cal G}_\nu\ \right)\ <\ \delta,
\nn\ee
  then for all $t\ne0$  
\be
\rho\left(\ \vec\psi(t)\ ,\ {\cal G}_\nu \right)<\varepsilon.
\nn\ee
\end{theo}

\section{ Excitation thresholds for DNLS}

For a fixed lattice dimension, $d$, we consider the family of equations
(\ref{eq:dnls}) parametrized by $\sigma$. Theorem 2.1 gives a criterion
for the existence of a ground state. 
\bigskip

\nit {\bf Definition:}
 If for any $\nu>0$ the variational
problem (\ref{eq:Inudef}) has a strictly negative infimum, 
${\cal I}_\nu<0$ then, by Theorem 2.1, a ground state exists for any $\nu>0$. In this case we say
that there is {\it no excitation threshold}. However, if there is a
strictly 
 positive constant $\nu^D_{thresh}$ (which may depend on $d$ and $\sigma$) such
that ${\cal I}_\nu<0$ 
 if and only if $\nu>\nu^D_{thresh}$, then we call $\nu^D_{thresh}$ an
{\it excitation threshold} or $L^2$ {\it excitation threshold} for a ground state.
\bigskip

The main result concerning DNLS is the following:
\begin{theo}

\nit (1)\ Let $0<\ \sigma\ <\ {2\over d}$. Then, ${\cal
I}_\nu < 0$ for all $\nu>0$. Therefore, the variational problem
(\ref{eq:Inudef}) has a solution for all $\nu>0$ and there is no
excitation threshold.

\nit (2)\ Let $\sigma\ \ge\ {2\over d} $. Then, there exists a ground state
excitation threshold, $\nu^D_{thresh}>0$. 

\end{theo}
\medskip

\nit{\bf Remark on DNLS vs. NLS:}  Here we contrast the discrete equation,
DNLS, 
and its continuum limit. In particular, we comment on some consequences
of the breaking of various symmetries in passing from NLS to DNLS.

\nit (1) The continuum limit of (\ref{eq:dnls}) ($\kappa=h^{-2}$, $h\ =\ $
lattice spacing, and $h\to0$) is the $d-$dimensional
nonlinear Schr\"odinger equation:
\be
i\D_t\phi\ =\ -\Delta\ \phi\ -\ |\phi|^{2\sigma}\phi,\label{eq:nls}
\ee
For initial data $\phi(t=0,x)\in H^1(\R^d)$, it has been shown that there exists 
 a local solution
which is continuous in time with values in $H^1(\R^d)$ and which satisfies
 the analogous conservation laws \cite{kn:GV}, \cite{kn:Kato}. 
 If $\sigma < {2\over d}$
solutions are always global in time, while for $\sigma\ge {2\over d}$, finite energy
initial data may give rise to a solution which leaves the space $H^1(\R^d)$ after a
finite time \cite{kn:Glassey}, \cite{kn:VPT}, \cite{kn:nls83}. 
 In contrast, the evolution
for (\ref{eq:dnls}) is globally defined in time. 

\nit (2) Solitary standing waves can be found by methods analogous to those
used in section 2. An excitation threshold for standing waves, 
 in terms of the natural $L^2$
invariant:
\be
{\cal N}_{NLS}[\ \phi\ ]\ =\ \int_{\R^d}\ |\phi(x)|^2\ dx\nn
\ee 
exists only in the case $\sigma={2\over d}$. This 
 follows because under the scaling: 
\be
\phi(x,t)\mapsto \phi_\rho(x,t)\equiv
\rho^{1\over\sigma}\phi(\rho x,\rho^2 t),\label{eq:scaling}\ee
 we find 
 \be
 {\cal N}[\ \phi_\rho\ ]\ =\ \rho^{{2\over\sigma}-d}\ {\cal N}[\ \phi\ ].
 \label{eq:scaling-property}\ee
Thus, given that a single standing wave exists, 
if $\sigma\ne{2\over d}$, 
scaling can be used to find one of arbitrarily small total
power, ${\cal N}_{NLS}$. In contrast, the dilation
symmetry is broken in the discrete case.  

\nit (3) Let $\sigma={2\over d}$,  and let $R$ denote the ground state
standing wave. That is, $R$ is an $H^1$ solution of 
 $\Delta\ R\ -\ R\ + \ R^{{4\over d}+1}=0$
 of minimal power ${\cal N}_{NLS}\ \equiv\ {\cal N}_{thresh}$. 
 In \cite{kn:nls83}, \cite{kn:survey} it was proved that if 
\be {\cal N}[\ \phi_0\ ]\ < \ {\cal N}_{thresh}\nn\ee
then the solution exists for all time and disperses to zero in the sense
that $\| \phi(t) \|_{L^p}\ \to\ 0$, as $|t|\to\infty$, for $p>2$. 

\nit {\bf Conjecture:} {\it If ${\cal N}[\ \vec\psi_0\ ]\ <\ \nu^D_{thresh}$, then 
the solution of DNLS disperses to zero in the sense that for any $p\in (2,\infty]$:
\be \|\vec\psi(t)\|_{l^p(\Z^d)}\ \to\ 0,\ {\rm as}\ |t|\to\infty.\nn\ee
}

\nit Proofs of the assertions in  (3) are
 given in \cite{kn:survey},\cite{kn:nls83} and 
 rely on the pseudo-conformal
symmetry of the continuum limit NLS, a symmetry which is absent in DNLS.
\bigskip

Theorem 3.1 is a consequence of Propositions 4.1 and 4.2 of section 4. We begin
by investigating the conditions on
$\sigma,\ d$, and  $\nu$  under which ${\cal I}_\nu<0$.
\medskip

\begin{prop}
${\cal I}_\nu\ \ge\ 0$ if and only if $\nu$ is such that the following
inequality holds for all $\vec u\in l^2(\Z^d)$:
\be
\sum_{l\in\Z^d}\ |u_l|^{2\sigma+2}\ \le\ (\sigma+1)\ \kappa\ 
 \nu^{-\sigma}\ \left( \sum_{l\in\Z^d}\ |u_l|^{2}\right)^\sigma\ 
  \left(-\delta^2\vec u,\vec u\right).
  \label{eq:dsngnu}\ee
\end{prop}
\bigskip
To prove Proposition 3.1, we observe that ${\cal I}_\nu\ge0$ if and only if for all
$\vec u\in l^2(\Z^d)$, with $\|\vec u\|^2_{l^2}=\nu$
\be
(\sigma +1)^{-1}\ \sum_{l\in\Z^d} |u_l|^{2\sigma +2}\ \le\
\kappa \left( -\delta^2\vec u,\vec u \right).
\label{eq:xx}\ee
Let $\vec 0\ne\ \vec v\in l^2$ be arbitrary. Then, if $\vec u$ defined by:
\be
\vec u\ \equiv\ \sqrt{\nu}\ \| \vec v\ \|_{l^2}^{-1}\ \vec v\nn
\ee
satisfies the inequality (\ref{eq:xx}), which after some algebra yields
(\ref{eq:dsngnu}). Finally, if ${\cal I}_\nu\ \ge\ 0$ we have that  ${\cal I}_\nu\ =\
0$. This is seen by simply taking a sequence whose $N^{th}$ element is a constant
(depending on $\nu$) on the set of sites satisfying $|l|\le N$ and zero
otherwise. Along 
 such a sequence we have ${\cal N}\ =\ \nu$ and ${\cal H}$ tending to
zero. Therefore, ${\cal I}_\nu\ =\ 0$.

\bigskip

\nit{\bf Strategy of the proof of Theorem 3.1:}
 Clearly, if the inequality (\ref{eq:dsngnu}) holds for
some $\nu_1$ then it holds for all $\nu\le\nu_1$.  We shall prove in
Proposition 4.2d that a ground
state does not exist for any $0\le\nu\le\nu_1$. We are interested in
characterizing 
$\nu_{thresh}$  defined by:
\be
\nu^D_{thresh} \ \equiv\ \sup\{\ \nu:\ {\rm inequality\ (\ref{eq:dsngnu})\  holds } \}.
\nn\ee

In the following section we 
relate this threshold value to the  
optimal (best) constant in an interpolation estimate related to the Hamiltonian energy.
${\cal H}$. 
If a finite positive $\nu^D_{thresh}$ exists, then for any $\nu>\nu^D_{thresh}$
 and element of $l^2(\Z^d)$, 
 $\vec u_*$,   
 can be found which violates the inequality (\ref{eq:dsngnu}).
 This choice of $\vec u_*$ shows that ${\cal I}_\nu<0$, and by Theorem
 2.1 there is a
 ground state. If, however, for any choice of $\nu>0$ 
  one can construct an element of $\l^2$ for which  the
   inequality (\ref{eq:dsngnu}) is violated, Theorems 3.3 and 2.1 imply that a ground
   state exists for any $\nu>0$, {\it i.e.} there is no excitation threshold.
  The strategy used to prove Theorem 3.1 is to show that if
  $0\ <\ \sigma\ <\ {2\over d}$, then there is no value of $\nu$ for which the inequality 
   (\ref{eq:dsngnu}) 
  holds for arbitrary $\vec u\in\ l^2$. However, if
   $\sigma\ge {2\over d}$ we show it holds
  if and only if $\nu\le\nu^D_{thresh}$,  for some $\nu^D_{thresh}>0$.
 \bigskip

\section{Best constants and excitation thresholds for DNLS}

In this section we relate the problem of 
 characterizing  excitation thresholds to the problem of finding the optimal or
 best constant in discrete interpolation inequalities of
 Sobolev-Nirenberg-Gagliardo type.

The discussion concluding section 3 motivates the following question, answered in
Theorem 4.1 below: 

\nit  {\it When does there exist a constant $C>0$ such that for all} 
$\vec u\ =\ \{u_l\}\ \in\ l^2(\Z^d)$:
\be
\sum_{l\in\Z^d}\ |u_l|^{2\sigma+2}\ \le\ C\ 
 \left(\ \sum_{l\in\Z^d}\ |u_l|^2\ \right)^\sigma\ \left(\ -\delta^2\vec u,\vec
 u\ \right)?
 \label{eq:dsng}
 \ee
\medskip

If (\ref{eq:dsng}) holds for some $C>0$ and $C_*$ is the infimum over all such
constants, then $\nu^D_{thresh}$ defined by 
\be
(\sigma +1)\ \kappa\ \left(\nu_{thresh}^D\right)^{-\sigma}\ \equiv\ C_*
\label{eq:nuthreshdef}
\ee 
is a ground state excitation threshold. Therefore, we seek to characterize the
optimal constant, $C_*$. If there is a strictly positive and finite $C_*$, then
\be
{1\over{C_*}}\ =\ {\cal J}^{\sigma, d}\ \equiv\ 
\inf\ {\left(\ \sum_{l\in\Z^d}\ |u_l|^2\ \right)^\sigma\ 
 \left(-\delta^2\vec u,\vec
 u\right)\over \sum_{l\in\Z^d}\ |u_l|^{2\sigma+2}}
\label{eq:Jdef}
\ee
\medskip
and we have:
\be 
\nu^D_{thresh}\ =\ \left(\ (\sigma + 1)\ \kappa\ {\cal J}^{\sigma, d}\
\right)^{1\over\sigma}.
\label{eq:nuthresh}
\ee
\medskip

\nit {\bf Remark:} 
 If ${\cal J}^{\sigma,d}>0$, then by Proposition 4.2 below,
  there exists a strictly positive lower bound on the
energy, ${\cal N}$, of a ground state.

\medskip

Note that (\ref{eq:nuthresh}) is consistent with the simple observation that for
the case of uncoupled lattice sites, $\kappa=0$, there is no excitation
threshold. For example, in this case the solution 
\ba \psi_0(t)\ &=&\ \nu^{1\over2}\ e^{i|\nu|^{\sigma}t}\nn\\
    \psi_l(t)\ &=&\ 0\ \ l\ne0\ \label{eq:onesite}
\ea
is a $l^2(\Z^d)$ solution of (\ref{eq:dnls}) with ${\cal N}=\nu$.  This limit is also called the
{\it anti-integrable limit} \cite{kn:MA}, \cite{kn:Aubry}. In section 5 
we shall relate the anti-integrable limit to the large amplitude limit of our variationally
constructed ground states.

\begin{prop}
  If $\sigma< {2\over d}$, then ${\cal J}^{\sigma, d}=0$. Therefore, for
$\sigma< {2\over d}$, and there is no ground state
excitation
threshold ($\nu^D_{thresh}=0$). In other words, ground states of arbitrary energy, ${\cal
N}$, exist.
\end{prop}

\nit{\it proof of Proposition 4.1:} 
 Consider the one parameter family of trial functions,
$\vec u(\alpha)$ defined by:
\be
u_l(\alpha)\ =\ e^{-\alpha\ |l|},
\label{eq:trial}
\ee
where $l=(l_1,...,l_d)\in\Z^d$, 
 $|l|\ =\ |l_1|+...+|l_d|$ and $\alpha>0$.
 Evaluation of the terms of the quotient in (\ref{eq:Jdef}) yields, for
 $\alpha\downarrow 0$:
 \be
 \sum_{l\in\Z^d}\ |u_l|^2\ \sim\
\alpha^{-d},\ \  
\left(-\delta^2\vec u,\vec u\right)\ \sim\ \alpha^{2-d},\ \  
 \sum_{l\in\Z^d}\ |u_l|^{2\sigma+2}\ \sim\ \alpha^{-d}.
 \nn\ee
 Therefore, the quotient in (\ref{eq:Jdef}) is of order
 $\alpha^{2-d\sigma}$, which tends to zero as $\alpha$ tends to zero if
 $\sigma < {2\over d}$. This proves the Proposition 4.1.
\medskip

\begin{prop} Let $\sigma\ge{2\over d}$. Then,

\smallskip

\nit (a)  ${\cal J}^{\sigma, d}>0$.

\nit (b) If $\|\ \vec\psi\ \|_{l^2}^2\ =\ \nu$, then
\be
{\cal H}_D[\ \vec\psi\ ]\ \ge\ \kappa\left(-\delta^2\vec\psi,\vec\psi\right)\
\left[1-\left({\nu\over\nu^D_{thresh}}\right)^\sigma\right],
\label{eq:HDlb}
\ee
where $\nu^D_{thresh}>0$ is given by (\ref{eq:nuthresh}). 
\nit For $\sigma\ge {2\over d}$,\   
 $\nu^D_{thresh}(\sigma,d)$ is an excitation threshold, {\it i.e.}\  

\nit (c) if  $\nu>\nu^D_{thresh}(\sigma,d)$ then ${\cal I}_\nu<0$ and a ground state
exists, and

\nit (d)  if $\nu<\nu^D_{thresh}(\sigma,d)$, then $I_\nu=0$ and there is no ground state
minimizer of (\ref{eq:Inudef}).
\end{prop}
\medskip

\nit{\it proof of Proposition 4.2 :}
  To prove part (a) suffices to show that the inequality (\ref{eq:dsng})
holds for {\it some} positive constant, $C$. Then part (c) follows from the discussion at the
end of section 3. We proceed as follows. For
functions $f\in H^1(\R^n)$, one has the Sobolev-Nirenberg-Gagliardo
inequality \cite{kn:Friedman}:
\be
\|f\|_{2\sigma +2}^{2\sigma+2}\ \le C\ \|\nabla f\|_2^{\sigma n}\
 \| f\|_2^{2+\sigma (2-n)},\label{eq:sng}
 \ee
 where $\sigma$ is restricted to satisfy:
 \ba 0\ &<&\ \sigma\ <\ \infty,\ \ n=1,2\nn\\
	 0\ &<&\ \sigma\ <\ 2(n-2)^{-1},\ n\ge3.\label{eq:sigmarestrict}
 \ea
The proof of (\ref{eq:sng}) can be followed closely to yield, under the same
restrictions on $\sigma$, the following estimate in the discrete case
for $\vec u\in\ l^2(\Z^d)$:
\be
\sum_{l\in\Z^d}\ |u_l|^{2\sigma+2}\ \le\ C\
 \left(\ \sum_{l\in\Z^d}\ |u_l|^2\ \right)^{1+{\sigma\over2}(d-2)} \
  \left(\ -\delta^2\vec u,\vec u\ \right)^{\sigma d\over2}.	
\label{eq:sng2}
\ee
To give the idea, we present the proof of (\ref{eq:sng2}) in the case $d=2$.
We write $u_l=u_{ab},\ (a,b)\in\Z^2$. Without loss of generality we can
take $u_{ab}\ge0$. 
Note that
\be
u_{ab}^{\sigma+1}\ =\ \sum_{\a =-\infty}^a\ \left(\ u_{\a b}^{\sigma+1}\ -\
u_{\a -1,b}^{\sigma+1}\ \right).\nn
\ee
By the fundamental theorem of calculus, 
\ba
  u_{\a b}^{\sigma+1}\ -\
 u_{\a -1,b}^{\sigma+1}\ &=&\ \int_0^1\ {d\over ds}\left[\ 
  su_{\a b}\ +\ (1-s)u_{\a -1,b}\ \right]^{\sigma +1}\ ds\nn\\
  &=&\ (\sigma +1)\ \int_0^1\ \left[\ su_{\a b}\ +\ (1-s)u_{\a -1,b}\
  \right]^\sigma\ ds\ \left(\ u_{\a b}\ -\ u_{\a -1,b}\ \right)\nn\ea
Therefore, (using the convention that sums without specified upper and lower limits 
 are understood to be taken over all $\Z^d$)
\be
\left|\ u_{\a b}^{\sigma+1}\ -\ u_{\a -1,b}^{\sigma+1}\ \right|\ \le\ 
 |\sigma +1|\ \sum_\a\ 
  \max\left( |u_{\a b}|^\sigma\ ,\ |u_{\a -1 ,b}|^\sigma \right)
   \left|\ u_{\a b}\ -\ u_{\a -1,b}\ \right|.
   \nn\ee
It follows by summing over $\alpha$ and applying the Cauchy-Schwarz inequality, that 
\be
\sum_\a\ \left|\ u_{\a b}^{\sigma+1}\ -\ u_{\a -1,b}^{\sigma+1}\ \right|\ \le
  2^{1\over2}|\sigma+1|\ 
  \left(\sum_\a\ |u_{\a b}|^{2\sigma}\right)^{1\over2}\ 
  \left(\sum_\a\ |u_{\a b}-u_{\a -1,b}|^2\right)^{1\over2}
  \nn\ee
The analogous computation can be performed by summing on the second index, to
get:
\be
\sum_\b\ \left|\ u_{a \b}^{\sigma+1}\ -\ u_{a ,\b -1}^{\sigma+1}\ \right|\ \le
  2^{1\over2}|\sigma+1|\  
	\left(\sum_\b\ |u_{a\b}|^{2\sigma}\right)^{1\over2}\ 
	  \left(\sum_\b\ |u_{a \b}-u_{a,\b -1}|^2\right)^{1\over2}.
		\nn\ee
The product of the last two estimates yields:
\ba
|u_{ab}|^{2\sigma+2}\ \le\ 2|\sigma +1|^2\ &&\left(\sum_\a\ |u_{\a
b}|^{2\sigma}\right)^{1\over2}\  
  \left(\sum_\a\ |u_{\a b}-u_{\a -1,b}|^2\right)^{1\over2}\nn\\
   \times\ &&\left(\sum_\b\ |u_{a\b}|^{2\sigma}\right)^{1\over2}\ 
		 \left(\sum_\b\ |u_{a\b}-u_{a,\b -1}|^2\right)^{1\over2}.
		 \nn\ea
Summing on $a$ and applying the Cauchy-Schwarz inequality gives:
\ba
\sum_a\ |u_{ab}|^{2\sigma +2}\ \le\ 
2|\sigma +1|^2\ &&\left(\sum_\a\ |u_{\a
b}|^{2\sigma}\right)^{1\over2}\  
  \left(\sum_\a\ |u_{\a b}-u_{\a -1,b}|^2\right)^{1\over2}\nn\\
 \times\ &&\left(\sum_{a,\b}\ |u_{a\b}|^{2\sigma}\right)^{1\over2}\ 
		   \left(\sum_{a,\b}\ |u_{a\b}-u_{a,\b -1}|^2\right)^{1\over2}.
					\nn\ea
Finally, summing this result on $b$ and applying the Cauchy-Schwarz inequality
gives (\ref{eq:sng2}) for the case $d=2$ and arbitrary $\sigma>0$.

To complete the proof of Proposition 4.2, we write 
 estimate (\ref{eq:sng2}) as:
\be
\sum_{l\in\Z^d}\ |u_l|^{2\sigma+2}\ \le\ C\
 \left(\ \sum_{l\in\Z^d}\ |u_l|^2\ \right)^\sigma\ 
 \left(\ -\delta^2\vec u,\vec u\ \right)\ 
\left(
{\left(\ -\delta^2\vec u,\vec u\ \right)\over \sum_{l\in\Z^d}\ |u_l|^2}
\right)^{{\sigma d\over2} -1}
\label{eq:sng3}\ee

The last factor in (\ref{eq:sng3}) is bounded by a constant for
$\sigma\ge{2\over d}$; the discrete Laplacian is a bounded operator. 
Therefore, if in addition to (\ref{eq:sigmarestrict}), we have
 $\sigma\ge{2\over d}$, then the estimate (\ref{eq:dsng})
holds. 

 Finally,  we want to show that for $d\ge3$ we can relax the constraint
 $0\ <\ 2-\sigma (d-2)$. Suppose $d\ge3$ and take $\sigma$ in the range for
 which we know the estimate (\ref{eq:dsng}) to hold. This estimate is equivalent
 to:
 \be
 \sum_{l\in\Z^d}\ |u_l|^{2\sigma+2}\ \le\ 
  C\ \left(\ -\delta^2\vec u,\vec u\ \right)
 \label{eq:dsng4}\ee
 subject where $\vec u$ satisfies the constraint
 \be
 \sum_{l\in\Z^d}\ |u_l|^2\ =\ 1.
 \label{eq:l2}
 \ee
 The constraint (\ref{eq:l2}) implies that for all $l\in\Z^d$, $|u_l|\le1$ and
 therefore if $\sigma_1$,  is {\it any} number satisfying
 $\sigma_1>\sigma\ge{2\over d}$, then the estimate (\ref{eq:dsng4}) holds with
 $\sigma$ replaced by $\sigma_1$. This implies the following result
 which completes the proof of part (a)  Proposition 4.2:
 \begin{theo}
For $\sigma\ge {2\over d}$, the interpolation inequality (\ref{eq:dsng})
holds.
 \end{theo}

 \nit{\bf Remark:} Note that there is no upper restriction for $n\ge 3$
 on $\sigma$
 as in the continuum case (\ref{eq:sng}). Through (\ref{eq:sng3}),  
 the boundedess of the discrete Laplacian, $-\delta^2$, on $l^2(\Z^d)$
 plays a key role.  

Part (b) of Proposition 4.2 follows from the definition of ${\cal H}_D$ and the
inequality 
(\ref{eq:dsngnu}) with optimal choice $\nu=\nu^D_{thresh}$ given by
(\ref{eq:nuthreshdef}).

Finally, we prove part (d) of Proposition 4.2. Suppose $\nu<\nu^D_{thresh}$. Then, by part
 (b), ${\cal
I}_\nu\ge0$. On the other hand, 
 as at the end of the proof of Proposition 3.1, we  have that ${\cal I}_\nu\le0$.
It follows that  $I_\nu=0$ for any $\nu<\nu^D_{thresh}$. 
 If the minimum is attained at a state $\vec
 \psi$, then
 \ba
 \kappa\ \left(\ -\delta^2\vec\psi\ ,\ \vec \psi\ \right) \ &=&\ {1\over{\sigma+1}}
\  \sum_l\ |\psi_l|^{2\sigma+2}\nn\\
\sum_l\ |\psi_l|^2\ &=&\ \nu\nn\ea 
  Since $\sigma\ge{2\over d}$, $\nu^D_{thresh}$ defined by (\ref{eq:nuthresh}) is
  strictly positive and by (\ref{eq:dsngnu}), with the optimal choice 
   $\nu=\nu^D_{thresh}$,  
  we have
  \be
   \kappa\ \left(\ -\delta^2\vec\psi\ ,\ \vec \psi\ \right) \ \le
  \ \kappa\left(\nu\over\nu^D_{thresh}\right)^\sigma\ \left(\ -\delta^2\vec\psi\ ,\ \vec
  \psi\ \right)\ <\ \kappa \ \left(\ -\delta^2\vec\psi\ ,\ \vec
	\psi\ \right),\nn\ee
   a contradiction.

  Theorem 3.1 now follows from Propositions 4.1 and 4.2.

\section{Large amplitude and  the anti-integrable limit}

\centerline{\it Large $\nu$  limit of ground states}

As discussed in \cite{kn:MA}, 
breather solutions of DNLS can also be constructed perturbatively in 
the limit of zero coupling, $\kappa\equiv0$, also called the {\it
anti-integrable limit}; see also \cite{kn:Aubry}. In \cite{kn:MA}, as an explanation
for the numerical studies in \cite{kn:BRC}, it is conjectured
that the large amplitude {\it anti-integrable limit breathers} play an important 
role in the
dynamics of DNLS. We now give evidence of this, by showing the connection
between the nonlinearly stable ground state breathers constructed by variational
methods and the large amplitude {\it anti-integrable breathers}. 
We also prove that as $\nu$ increases,  
 ground state breathers of "total power" $\nu$ grow in amplitude and become  
increasingly concentrated about one lattice site. This property of ground states
and their nonlinear stability (Theorem 2.2) elucidate the numerical simulations
in \cite{kn:BRC}.

We begin by considering a scaled version of the variational problem
(\ref{eq:Inudef}), for the DNLS ground state:
\be
{\cal I}_\nu\ =\
  \inf\{\ -{\kappa}
	\left( \delta^2 \vec f,\ \vec f \right)\ -\
	  {1\over{\sigma +1}}\ \sum_l\ |f_l|^{2\sigma +2}\ :\ \sum_l\
	  |f_l|^2\ =\ \nu\
		\}.
		  \label{eq:dnlsgs}
			\ee
 In anticipation of our  taking $\nu\uparrow\infty$, we set
\be
f_l\ =\ \nu^{1\over2}\ F_l,\ \ \sum_l\ |F_l|^2\ =\ 1.
\label{eq:phidef}
\ee
and  introduce the parameter
\be \alpha\ =\ \alpha(\nu,\kappa;\sigma)\ \equiv
 {\kappa\over\nu^\sigma}, \label{eq:alpha}\ee
 which tends to zero as $\nu$ tends to infinity. 
The variational problem, (\ref{eq:dnlsgs}), is then equivalent to:
\be
{\cal K}({\alpha})\ =\ 
  \inf\{\ {\alpha}\ 
  \left( -\delta^2 \vec F,\ \vec F \right)\ -\ 
  {1\over{\sigma +1}}\ \sum_l\ |F_l|^{2\sigma +2}\ :\ \sum_l\ |F_l|^2\ =\ 1\
  \}.
  \label{eq:scaledenergy}
  \ee
  By Theorem 3.1, if $\nu\ge\nu^D_{thresh}\ge0$, then there is a ground state
  breather:
  \be
  \vec G\ =\ \vec G({\alpha})\ =\ \{G_l(\alpha)\}_{l\in\Z^d},
  \label{eq:Ggs}
  \ee
  satisfying the Euler-Lagrange equation:
  \be
  -{\alpha}  \left( \delta^2\ G \right)_l
  \ - |G_{l}|^{2\sigma}\ G_{l}\ =\ \lambda_\alpha\ G_{l},
  \label{eq:anti}\ee
where $\lambda_\alpha$ is a Lagrange multiplier.
By (\ref{eq:phidef}), this gives rise to a ground state family of solutions of
(\ref{eq:dnlsgs}): 
\be
\vec\psi_g(t)\ =\ \nu^{1\over2}\ \vec G(\alpha)\ e^{-i\lambda_\alpha\nu^\sigma t}\
e^{i\gamma},\ \gamma\in [0,2\pi).
\nn\ee
By Theorem 2.2, the ground state family is  nonlinearly orbitally  stable, and
is therefore expected to participate in the dynamics.

  {\it What is the structure of ground states for large $\nu$? We next
  show that as $\nu\to\infty$, ground states become concentrated on the
  lattice about a single site.}

  To see this, we first observe that by the methods of appendix A (see
  also \cite{kn:colinweinstein}), 
  
  \nit (a) As 
 $\alpha$ tends to zero ($\nu\uparrow\infty$) through a sequence,
 $\{\vec G(\alpha)\}$, 
 is a minimizing sequence for the limit variational problem:
\be
 {\cal K}_\infty\ =\
  \inf\{\
  -{1\over{\sigma +1}}\ \sum_l\ |G_l|^{2\sigma +2}\ :\ \sum_l\ |G_l|^2\ =\
   1\ \}.
		 \label{eq:limitingscaledenergy}
\ee

\nit (b) A subsequence can be extracted, which (modulo phase adjustments)
 converges to a minimizer, $\vec
G^\infty$. $ \vec G^\infty$ satisfies the Euler-Lagrange equation associated with the
limit problem (\ref{eq:limitingscaledenergy}):
\ba
 - |G_{l}^\infty|^{2\sigma}\ G_{l}^\infty\ &=&\ 
  \lambda_0 
  \ G_{l}^\infty,\nn\\ 
 \sum_l\ |G^\infty_{l}|^2\ &=&\ 1.
\label{eq:eleqn}
\ea

Thus, for each $l\in\Z^d$, 
 $G_{l}^\infty\in\{0\}\cup\{(-\lambda)^{1\over {2\sigma}}\ e^{i\gamma}\ :\
 \gamma\in [0,2\pi)\ \}$.
Since $\|\ \vec G^\infty\ \|_{l^2(\Z^d)} = 1$, $G_l^\infty$ 
can be nonzero only at a finite
number of sites, $N\ge1$. Therefore,
\ba
\|\ \vec G^\infty\ \|_{l^2(\Z^d)} &=& 1\ \ {\rm implies\ }\ -\lambda\ =\ N^{-\sigma}\nn\\
-{1\over{\sigma +1}}\ \sum_l\ |G_l^\infty |^{2\sigma+2}\ &=&\ -{N^{-\sigma}\over{\sigma+1}}.
\nn\ea
The minimum is therefore attained for $N=1$ and we have:
$\vec G_{l}^\infty\ =\ \pm\delta_{ll_0}$ for some $\l_0\in\Z^d$, and
$\lambda_0=-1.$

Therefore, as $\nu\to\infty$, a subsequence of ground states converges,  
to a limiting state:
\be
G_{l}(\alpha)\ \to\ G_l^\infty\ \equiv\  \delta_{l,l_0},\ {\rm in}\ l^2(\Z^d)
\nn
\ee
for some $l_0\in\Z^d$.
Therefore the large $\nu$ ($\alpha$ small) 
 limit of ground states behaves as a  large amplitude 
 {\it one-site breather}:
\be
\psi_l(t)\ \sim\ 
 \pm\nu^{1\over2}\ \delta_{l,l_0}\ e^{-i\nu^\sigma t},\ \ {\rm for\ \nu\ large}.
\label{eq:onesitebreather}
\ee

\centerline{\it Connection with the anti-integrable limit}

For small $\alpha$ equation (\ref{eq:anti}) is  the anti-integrable
limit studied in \cite{kn:MA}. The approach taken in \cite{kn:MA},
\cite{kn:Aubry} is to first observe that for $\alpha=0$ each lattice
site evolves independently and that (\ref{eq:anti})
has solutions $\Psi_l(t)$, where for each $l\in\Z^d$, $\Psi_l$ satisfies
the equation:
\be
i\D_t\Psi(t)\ =\ -|\Psi(t)|^{2\sigma}\ \Psi(t).\label{eq:kappa0}
\ee
The solutions of (\ref{eq:kappa0}) are:
\be \Psi(t)\ =\ \omega\ e^{i|\omega|^{2\sigma} t}\ e^{i\gamma},
\label{eq:oscillations}\ee
with  $\omega,\gamma\in\R$.
Fix a solution which is supported at 
 lattice sites $q\in I\subset\Z^d$, where $I$ is finite or infinite and
  such that at each site
 the evolution is an oscillation of the form 
 \be \Psi_q(t)\ =\ \omega_q\ e^{i|\omega_q|^{2\sigma} t}\ e^{i\gamma_q}, 
 \ q\in I
 \nn\ee
 and such that the frequencies $\omega_q$ are all commensurate. 
The implicit function theorem implies that these solutions 
have a continuation  for $\alpha$ 
sufficiently small in the space of time-periodic solutions.
These range in spatial complexity from those that are  
small perturbations of the simplest $\alpha=0$ breather, consisting of a
solution of the form (\ref{eq:onesite}), to those which are small perturbations of an
 $\alpha=0$ "spatially chaotic" configuration of oscillators.

Consider the continuation from the anti-integrable limit of one-site breathers.
These  are solutions of the form:
\be \Psi_l(t,\alpha,\mu)\ =\ A_l(\alpha,\mu)e^{-i\mu t},\ l\in \Z^d\nn\ee
where
\ba
\mu A_l\ &=&\ \alpha\left(\delta^2A\right)_l\ -\ |A_l|^{2\sigma}A_l,\nn\\
A_l(\alpha=0,\mu)\ &=&\ (-\mu)^{1\over 2\sigma}\delta_{l,l_0},\ {\rm for\ some}\ l_0\in\Z^d.
\label{eq:Aeqn}
\ea
We wish to relate the two-parameter family 
$\vec A(\alpha,\mu)$ to the family of
scaled ground states $\vec G(\alpha)$, for small $\alpha$. Note that 
 $\vec A(0,-1)$ is
such that $\|\vec A(0,-1)\|_{l^2(\Z^d)}=1$.  It is easy to check, by the implicit function
theorem that a locally unique solution
 $\left(\alpha,\mu(\alpha)\right)$ defined in
a neighborhood of $\alpha=0$ exists  such that 
\ba\|\vec A(\alpha,\mu(\alpha))\|_{l^2(\Z^d)}\ &=&\ 1,\nn\\
	\mu(0)\ =\ -1.\nn
	\ea
Therefore, by our variational arguments and local uniqueness:
\be
\vec G(\alpha)\ =\ \vec A\left(\alpha,\mu(\alpha)\right).
\nn\ee

\section{Thresholds for coupled systems of nonlinear Schr\"odinger equations}

In this section we discuss results for systems of coupled on nonlinear
Schr\"odinger equations (CNLS) ({\ref{eq:cnls}):
\ba
i\D_t\ \psi_l\ +   \ \D_x^2\psi_l\ +\
  \kappa\ 
  \left(\ \delta^2\ \vec\psi\ \right)_l\ 
  +\ (\sigma+1)|\psi_l|^{2\sigma}\ \psi_l\ &=&\ 0,
 \label{eq:cnls1}\\
 \vec\psi\ =\ \{\psi_l(t,x)\}_{l\in\Z^d},\ d=1,2,\ (t,x)&\in&\R^2\nn
 \ea
 $CNLS$ has been introduced as a model  governing the propagation of light pulses in
 a coupled $d=1$ or $d=2$ dimensional array of 
 optical fibers. We consider the case where the discrete variable varies
 over $\Z^d$, and such that $\psi_l(t,x)$ decays as $l$ and $x$ tend to infinity.
 Other boundary conditions ({\it e.g.} periodic are considered in \cite{kn:yearythesis},
 \cite{kn:BA}., \cite{kn:BA}.
We follow a similar outline for the CNLS as that followed in our analysis of
DNLS. Certain details are omitted and for them we refer to \cite{kn:WY},
\cite{kn:yearythesis}.

Given initial data $\vec \psi_0(x)$ for CNLS satisfying 
\be
\sum_{l\in\Z^d}\ \|\psi_{0l}\|_{H^1}^2\ <\ \infty,\nn
\ee
there is a unique solution $t\mapsto \vec \psi(t,x)$ which is continuous in $t$
with values in $l^2(\Z^d)\times H^1(\R)$. 
The following two functionals, evaluated on solutions, are  independent in time:
\ba
{\cal H}[\ \vec\psi\ ]\ &=&\ \int_\R\ \left( -\delta^2\vec\psi(x),\vec\psi(x) \right)
\ +\ \sum_{l\in\Z^d}\ \int_\R\ |\D_x\psi_l(x)|^2\ dx\ -\ 
 \int_\R\ |\psi_(x)|^{2\sigma+2}\ dx\nn\\
{\cal N}[\ \vec\psi\ ]\ &=&\ \sum_l \int_\R\ |\psi_l|^2\ dx.\nn\ea

${\cal H}$ is a Hamiltonian energy of the CNLS in the sense that CNLS can be
expressed as:
\be
i\D_t\vec\psi\ =\ {\delta{\cal H}\over\delta\vec\psi^*}\nn\ee
The functional ${\cal N}$ corresponds to the {\it total input power} in the system.

Of interest are nonlinear bound states of CNLS. These are solutions of the form:
\be
\vec\psi\ =\ e^{i\lambda^2 t}\ \vec g(x;\lambda),
\label{eq:boundstate1}
\ee
for which the invariants ${\cal H}$ and ${\cal N}$ are finite. The components of
$\vec\psi$ satisfy the coupled system of equations:
\be
-\lambda^2\psi_l\ +\ \D_x^2\psi_l\ +\ \kappa\ \left(\ \delta^2\vec\psi\
 \right)_l\ +\  (\sigma+1)|\psi_l|^{2\sigma}
\ \psi\ =\ 0\ l\in\Z^d.
\label{eq:gstate}
\ee

In analogy with the discrete case, we seek to characterize the ground state of the system by
variational methods.

\nit{\bf Definition:} Let

\be {\cal J}_\nu\ =\ \inf\{\ {\cal H}\left[\ \vec f\ \right]\ :\ 
 {\cal H}\left[\ \vec f\ \right]\ =\
\nu\ \}.
\label{eq:Jnudef}
\ee
\medskip
Because of the similarity of the arguments to those in the
previous sections and the more detailed treatment in
\cite{kn:WY},\cite{kn:yearythesis} we provide a summary.

\bigskip
\nit (1)  If $0<\sigma<2$, then  ${\cal J}_\nu\ >\ -\infty$ for any $\nu>0$.

\nit (2) In analogy with Theorem 2.1, we can show:
\begin{theo}
 The infimum in (\ref{eq:Jnudef}) is attained if and only if ${\cal J}_\nu<0$.
Moreover, any minimizing sequence has a subsequence which converges strongly in
$l^2\left(\Z^d;H^1(\R)\right)$ modulo translations in space and phase. Furthermore, 
any  minimizer satisfies the equation (\ref{eq:gstate}).
\end{theo}

In view of this result, we study the question: for which $\sigma, d$ and $\nu$ do we
have ${\cal J}_\nu<0$?
\medskip

\nit (3) \begin{prop} ${\cal J}_\nu\ =\ 0$ if and only if for any 
 $\vec\psi\in l^2\left(\Z^d;H^1(\R)\right)$ we have the estimate:
\be
\|\ \vec\psi\ \|_{2\sigma+2}^{2\sigma+2}\ \le\
\nu^{-\sigma}a_\sigma^{{\sigma\over2}-1}\ \|\ \vec\psi\ \|_2^{2\sigma}\
\langle-\delta^2\vec\psi\ ,\ \vec\psi\ \rangle^{1-{\sigma\over2}}\ \|\ \D_x\vec\psi\ \|_2^\sigma,
\label{eq:est3}
\ee
where $a_\sigma\ =\ \left({\sigma\over2}\right)^{\sigma\over
2-\sigma}-\left({\sigma\over2}\right)^{2\over 2-\sigma}.$
\end{prop}
Here, $\|\ \vec f\ \|_p^p\ =\ \left(\ \sum_i \|\ f_i\ \|_{L^p(\R)}^p\ \right)^{1\over p}$.

Proposition 6.1 is proved by a simple scaling argument. For any $\vec\psi$, such that 
$\|\ \vec\psi\ \|_2^2\ =\ \nu$, we define the scaling $\vec\psi^r(x)\ =\
r^{1\over2}\vec\psi(rx)$, which preserves the $L^2$ norm. Evaluation of the Hamiltonian on
$\vec\psi^r$ and minimization over $r>0$ gives:
\be
{\cal H}[\vec\psi^r]\ \ge {\cal H}[\vec\psi^{r_{min}}]\ =\ \langle -\delta^2\vec\psi\ ,\
\vec\psi\ \rangle\ -\ a_\sigma\ 
 \|\ \vec\psi\ \|_{2\sigma+2}^{ {4(1+\sigma)\over{2-\sigma}} }
 \| \D_x\vec\psi \|_2^{-{2\sigma\over{2-\sigma}} }\label{eq:Hlowerbound}
\ee
We can pass to an expression for arbitrary $\vec\psi$ by replacing $\vec\psi$ by 
$\nu^{1\over2}\|\ \vec\psi\ \|_2^{-1}\ \vec\psi$ in (\ref{eq:Hlowerbound}). 
This gives for {\it any} $\vec\psi\in l^2\left(\Z^d;H^1(\R)\right)$:
\be
{\cal H}[\vec\psi^{r_{min}}]\ =\ 
\left(\ {\nu\over\|\vec\psi\|_2^2} \right)\ 
  \langle -\delta^2\vec\psi\ ,\ \vec\psi\ \rangle\ -\ a_\sigma\
  \left(\ {\nu\over\|\vec\psi\|_2^2} \right)^{2+\sigma\over2-\sigma}
  \|\ \vec\psi\ \|_{2\sigma+2}^{ {4(1+\sigma)\over{2-\sigma}} }
   \| \D_x\vec\psi \|_2^{-{2\sigma\over{2-\sigma}} }\label{eq:Hlowerbound2}
   \ee
It follows that ${\cal J}_\nu$ can be realized as the infinimum of the expression in 
(\ref{eq:Hlowerbound2}) over all $\vec\psi\in l^2\left(\Z^d;H^1(\R)\right)$. 
 Thus ${\cal J}_\nu\ge0$ 
 if and only if (\ref{eq:est3}) holds for all $\vec\psi\in l^2\left(\Z^d;H^1(\R)\right)$.  
 As in the discrete case, it is simple to construct a sequence along which the $L^2$
 constraint is satisfied and the
 Hamiltonian tends to zero.

\nit (4) Suppose an estimate of the type (\ref{eq:est3}) holds. In particular, we let $C_*$
denote the smallest constant for which this estimate holds. That is,
\be
\|\ \vec\psi\ \|_{2\sigma+2}^{2\sigma+2}\ \le\
C_*\ \|\ \vec\psi\ \|_2^{2\sigma}\
\langle-\delta^2\vec\psi\ ,\ \vec\psi\ \rangle^{1-{\sigma\over2}}\ \|\ \D_x\vec\psi\
\|_2^\sigma,
\label{eq:est3best}
\ee
where
\be
C_*^{-1}\ =\ {\cal K}^{\sigma,d}\ \equiv\ 
\inf {\|\ \vec\psi\ \|_2^{2\sigma}\ \langle-\delta^2\vec\psi\ ,\ \vec\psi\
\rangle^{1-{\sigma\over2}}\ \|\ \D_x\vec\psi\ \|_2^\sigma\ \over
\|\ \vec\psi\ \|_{2\sigma+2}^{2\sigma+2}} \label{eq:est4}\ee

There are two possibilities. First, if ${\cal K}^{\sigma,d}\ =\ C_*^{-1}\ =\ 0$, then for any
$\nu>0$, there is
 a choice of $\vec\psi$ which makes the Hamiltonian negative. In this case, by assertion
 (2), a ground state of any prescribed $L^2$ norm exists; there is no $L^2$-  excitation threshold.
The second possibility is that $0\ <\ C_*^{-1}\ = {\cal K}^{\sigma,d}\ <\ \infty$. 
 In this case, 
we have that $J_\nu\ge0$ if and only if $C_*\ \le\ \nu^{-\sigma}\ a_\sigma^{{\sigma\over2}-1}$.
Therefore, we can define the {\it threshold power}, $\nu_c\ =\ \nu_c(\sigma,d)$ by:
\be
\nu_c\ =\  a_\sigma^{{1\over2} - {1\over\sigma}}\ C_*^{-{1\over\sigma}}\ =\
a_\sigma^{{1\over2} - {1\over\sigma}}\ \left({\cal K}^{\sigma,d}\right)^{1\over\sigma}.
\label{eq:nucdef}
\ee

Use of the estimate (\ref{eq:est3}) with the optimal choice $\nu=\nu_c$, we obtain the
sharp lower bound for the Hamiltonian, in analogy with the discrete
case (compare (\ref{eq:HDlb})):
\be
{\cal H}[\ \vec\psi\ ]\ \ge\ \langle -\delta^2\ \vec\psi\ ,\ \vec\psi\rangle\ \left(
\ 1\ -\ \left(\nu\over\nu_c\right)^\sigma\ \right).
\label{eq:Hlowerbound3}
\ee
for any $\vec\psi$ with $\|\ \vec\psi\ \|_2^2\ =\ \nu.$
\medskip

\nit (5) The question of when an $L^2$ threshold exists is
reduced to the determination of the range of values of $\sigma$ and $d$ for which  
${\cal K}^{\sigma,d}>0$. Formula (\ref{eq:nucdef}) then gives an expression for the threshold.
 To determine when ${\cal K}^{\sigma,d}$ is strictly positive amounts to determining 
  when one can prove an
 inequality of type (\ref{eq:est3best}) for some (not necessarily optimal) choice of $C_*$.
This is addressed in \cite{kn:WY}, \cite{kn:yearythesis}. Ranges of $\sigma , d$ for which
this inequality fails to hold are determined by scaling arguments, while a proof of such 
 inequalities for certain $\sigma, d$ can be obtained following the strategy used in the
 fully discrete case, where we mimic
  the proof of continuum interpolation estimates 
 ({\it e.g.} see the proof of Proposition 4.2) or alternatively by 
 applying the continuum interpolation estimates to functions of $d+1$ variables and where
 the functions are taken to be piecewise linear in the variable corresponding to the $d$
 discrete variables; see \cite{kn:WY}, \cite{kn:yearythesis}.
 The results
obtained are that ${\cal K}^{\sigma, d}>0$ for all $\sigma\in [1,2)$ and for all $\sigma\in
({2\over{d+1}}, {2\over d-1})$.
In summary we have:
\begin{theo} Let $\sigma\in [1,2)$ or $\sigma\in ({2\over d+1},{2\over d-1})$. Then,
there exists an $L^2$ excitation threshold given by $\nu_c$ in (\ref{eq:nucdef}).
\end{theo}
\bigskip

\centerline{\it Estimates on $\nu_c$ for $\sigma=1$, and $d=1$}
\medskip

We now consider the case $\sigma=1$ and $d=1$, an infinite one-dimensional array: 
\be
i\D_t\psi_n\ +\ \D_x^2\psi_n\ +\ \kappa\left(\ \psi_{n-1}\ -\ 2\psi_n\ +\ \psi_{n+1}\
\right)\ +\ 2\ |\psi_n|^2\ \psi_n\ =\ 0,\ n\in\Z \label{eq:1dcnls}
\ee
We show
how to get upper and lower estimates for the threshold power. A sketch was given in
\cite{kn:WY}, where an error appears in the displayed  upper and lower bounds (due to an
error in algebra).

By the above discussion, we know that there is an $L^2$ excitation threshold. That is, there
is a constant $\nu_c>0$ such that  there are no ground states of with $L^2$ norm less than
$\nu^{1\over2}$ and there are ground states of $L^2$ norm $\nu^{1\over2}$ for any
$\nu\ge\nu_c$. By (\ref{eq:nucdef}) (using that we must replace $\delta^2$ by
$\kappa\delta^2$ we have 
\be
\nu_c(1,1;\kappa)\ =\ 2 \kappa^{{1\over2}} {\cal K}^{1,1}\ =\ 
2 \kappa^{{1\over2}}\  \inf\ {\|\ \vec\psi\ \|_2^2\langle\ -\delta^2\vec\psi\ ,\ \vec\psi\
 \rangle^{1\over2}\ \|\ \D_x\vec\psi\ \|_2\over \|\ \vec\psi\ \|_4^4}.
\label{eq:nuc11estimate}
\ee
\medskip

\nit{\it Upper estimate on $\nu_c(1,1;\kappa)$:}

An upper estimate is obtained by evaluation of the functional in
 (\ref{eq:nuc11estimate})
on any $\vec\psi\ne0$. In particular, if we use as a trial function the exact {\it one-soliton}
supported on one site of the lattice, $\psi_j(x)={\rm sech}(x)\delta_{j0}$, we  obtain
the upper bound 
\be
\nu_c(1,1;\kappa)\ \le\ \kappa^{1\over2}\ 2\sqrt{6}\sim \kappa^{1\over2}\ 4.89...
\label{eq:upperbound}
\ee
\medskip

\nit{\it Lower estimate on }$\nu_c(1,1;\kappa)$:

To obtain a lower bound we follow the strategy in \cite{kn:WY}. First note
that   for arbitrary functions $\psi\in H^1(\R^2)$, 
\be
\| \psi \|_4^4\ \le\ C_{SNG}\ \left(\ \| \D_x\psi\ \|_2^2\
 +\ \|\ \D_y\psi\ \|_2^2\ \right)\ 
 \ \|\ \psi\ \|_2^2. 
 \label{eq:2dsng}
\ee
By scaling in $y$, $\psi(x,y)\mapsto\psi(x,ry)$, we  have from (\ref{eq:2dsng})
the estimate:
\be
\|\ \psi\ \|_4^4\ \le\ 2C_{SNG}\ \|\ \D_x\psi\ \|_2\
 \|\ \D_y\psi\ \|_2\ 
 \|\ \psi\ \|_2^3,
   \label{eq:SNG2}
	  \ee

In \cite{kn:nls83} the best constant in (\ref{eq:2dsng}) is calculated and was found to be:
\be
C_{SNG}\ =\ \left(\pi\ \times\ 1.86225...\right)^{-1}
\nn\ee
By (\ref{eq:nuc11estimate}), to obtain a lower bound for $\nu_c$ it sufficies to obtain an
lower bound for ${\cal K}^{1,1}$ or equivalently an upper bound for $C_*$. 

We next relate $C_*$ to $C_{SNG}$. This can be done by considering (\ref{eq:2dsng}) for the
restricted class  of functions, $\psi(x,y)$,
 which are smooth in $x$ and piecewise linear in $y$ with jumps in
$\D_y\psi(x,y)$ at the integers. 
 In particular, let
 \ba
 \psi(x,y)\ &=&\ (1-\theta)\ \psi_n(x)\ +\ \theta\ \psi_{n+1}(x),\nn\\
 y\ &=&\ n\ +\ \theta,\ \  0\le\theta\le1.\nn
 \label{eq:plpsi}
\ea
Direct calculation gives:
\ba
{2\over5}\ \sum_n\ \int\ |\psi_n(x)|^4\ dx\ &\le&\ \int\ |\psi(x,y)|^4\ dx\ dy\nn\\
 \sum_n\ \int\ |\psi_n(x)|^2\ dx\ &\ge&\ \int\ |\psi(x,y)|^2\ dx\ dy\nn\\
 \sum_n\ \int\ |\D_x\psi_n(x)|^2\ dx\ &\ge&\ \int\ |\D_x\psi(x,y)|^2\ dx\ dy\nn\\
 \sum_n\ \int\ |\ \psi_{n+1}(x)- \psi_{n}(x)\ |^2\ dx\ 
  &=&\ \int\ |\D_y\psi(x,y)|^2\ dx\ dy\nn\\
 \ea
This, together with (\ref{eq:SNG2}) yields:
\be
\|\ \vec\psi\ \|_4^4\ \le\ 5C_{SNG}\ 
 \| \vec\psi\ |_2^2\ \langle\ \vec\psi\ ,\ \vec\psi\ \rangle^{1\over2}\ \|\ \vec\psi\
 \|_2.\nn
\ee
with a non-optimal constant, $\tilde C=5C_{SNG}$ which is an {\it upper} bound for $C_*$.
Thus,
\ba
\nu_c(1,1,;\kappa)\ &\ge&\ 2\ \kappa^{1\over2}\ C_*^{-1}\ \ge\  2\ \kappa^{1\over2}\
(5C_{SNG})^{-1}\nn\\
 &=&\ \kappa^{1\over2}\ {2\over5}\pi\ \times\ 1.86225...\ \ge\ \kappa^{1\over2}\
2.3402...\label{eq:lowerbound}\ea

Combining (\ref{eq:upperbound}) and (\ref{eq:lowerbound}) we obtain:

\be
\kappa^{1\over2}\ 2.34...\ \le\  \nu_c(1,1,;\kappa)\ 
 \le\ \kappa^{1\over2}\ 4.89...
\label{eq:bounds}\ee
A careful numerical simulation \cite{kn:yearythesis} indicates $\kappa^{-{1\over2}}\
\nu_c(1,1;\kappa)\ \sim\ 4.08$.

\section{Appendix - Concentration Compactness Methods for DNLS}

Theorems 2.1 can be proved using the concentration compactness
principle; see, for example, \cite{kn:PLL}.
Since arguments follow quite closely those
for the continuum case (see \cite{kn:yearythesis}, \cite{kn:PLL} for a detailed
implementation), we present here an outline of the ideas.

Let $\vec u^{(k)}\ =\ \{u_l^{(k)}\}$,  
 denote a sequence in $l^2(\Z^d)$, and such
that 
\be \sum_l |u^{(k)}_l|^2\ =\ \nu.\nn\ee
Let $B_t(m)$ denote $\{l\in\Z^d\ :\ |l-m|<t\}$, and the norm $|l-m|\ =\
\max_{1\le i\le d} |l_i-m_i|$.

\begin{theo} (Concentration Compactness Principle)

There exists a subsequence $\vec u^{(n_k)}$ satisfying one of the following three
scenarios:
\medskip

\nit (1) Compactness ({\it the "mass" of the sequence concentrates}): 
 \ There exists $m_k\in\Z^d$ such that
for every  $\varepsilon>0$, there exists  a real positive number 
 $R_\varepsilon$ (independent of $k$), such that
\be
\sum_{l\in B_{R_\varepsilon}(m_k)}\ |u_l^{(n_k)}|^2\ \ge\ \nu - \varepsilon
\label{eq:compactness}\ee

\medskip

\nit (2) Vanishing ({\it the sequence spreads its mass over larger and larger
sets and tends to zero}): For all $R<\infty$,
\be
\lim_{k\to\infty}\ \sup_{m\in\Z^d}\ \sum_{l\in B_R(m)}\ |u_l^{(n_k)}|^2\ =\ 0
\label{eq:vanishing}
\ee

\medskip

\nit (3) Dichotomy ({\it the sequence concentrates its mass in at least two
 regions which become increasingly distant)}: There exists $\alpha\in (0,\nu)$ such that , for all $\varepsilon>0$,
there exist $k_0\ge1$ and disjointly supported 
 sequences $\vec a^{(k)},\ \vec b^{(k)}$ in $l^2(\Z^d)$ satisfying
for all $k\ge k_0$:
\ba
\|\ \vec u^{(n_k)}\ -\ \left(\vec a^{(k)} + \vec b^{(k)}\right)\ \|_{l^2(\Z^d)}\ &\le&
\varepsilon\nn\\
\left|\ \|\vec a^{(k)}\|_{l^2(\Z^d)}^2\ -\ \alpha\ \right|&\le& \varepsilon\nn\\
\left|\  \|\vec b^{(k)}\|_{l^2(\Z^d)}^2\ -\ (\nu-\alpha)\ \right|&\le& \varepsilon\nn\\
{\rm distance}\left({\rm supp}(a^{(k)}, {\rm supp}( b^{(k)}\right)\ &\to&\ \infty
\nn\ea
as $k\to\infty$.
\bigskip

\end{theo}

To prove this result, we introduce the sequence of  {\it concentration functions}:
 \be
  Q^{(k)}(t)\ =\ \sup_{m\in\Z^d}\ \sum_{l\in B_t(m)}\ |u_l^{(k)}|^2,
   \label{eq:Qdef}
	\ee
By following the arguments in \cite{kn:PLL} it can be shown that:

\nit (A) along a subsequence $n_k\to\infty$,  $Q^{n_k}(t)$ converges to a
nondecreasing and nonnegative function, $Q(t)$ with limit:
\be
\lim_{t\to\infty} Q(t)\ =\ \alpha\ \in\ (0,\nu).\nn\ee

\nit (2) the cases $\alpha=0$, $\alpha=\nu$ and $0<\alpha<\nu$  
 correspond, respectively, to the above scenarios: {\it vanishing, compactness} and {\it
 dichotomy}.

\bigskip

To prove Theorem 2.1 we must rule out the vanishing and dichotomy scenarios.

Vanishing is ruled out as follows. Let $\vec u_k$ denote a minimizing sequence. Then,
 ${\cal H}[\ \vec u_k\ ]\ =\ {\cal I}_\nu\ +\ \epsilon_k$, where
 $\epsilon_k\downarrow 0$ as $k\to\infty$. From the definition of ${\cal H}$ and the
 hypotheses ${\cal I}_\nu <0$ we have 
\ba
{\cal I}_\nu\ +\ \epsilon_k\ &=&\ {\cal H}[\ \vec u^{(k)}\ ]\nn\\
				&=&\ \langle\ -\delta^2\ \vec u^{(k)}\ ,\ \vec u^{(k)}\ 
				 \rangle\ -\ 
				 (\sigma+1)^{-1}\ \sum_{l\in\Z^d}\ | u_l^{(k)} |^{2\sigma+2}\nn\\
				 &\ge&\ -\ (\sigma+1)^{-1}\ \sum_{l\in\Z^d}\ 
				  | u^{(k)}_l |^{2\sigma+2}, 
\nn\ea
and therefore
\be
{(\sigma+1)\over2}\ |{\cal I}_\nu\ |\ \le\ \nu\ \|\ \vec u^{(k)}\ \|_{l^\infty}.
\label{eq:lb}\ee
Since vanishing implies $\|\ \vec u_k\ \|_{l^\infty}\to0$ as $k\to\infty$, the lower
bound (\ref{eq:lb}) precludes vanishing.

\bigskip
Dichotomy is ruled out, as in the continuum case \cite{kn:PLL},
 using the strict subadditivity of the functional ${\cal I}_\nu$,
 {\it i.e.}  if $0<\alpha<\nu$, then
\be
{\cal I}_\nu\  <\  {\cal I}_\alpha\ +\ {\cal I}_{\nu-\alpha}.
\label{eq:subadditivity}
\ee
The idea is as follows. If dichotomy occurs (see (3) above) then
as  $k\to\infty$ and $\varepsilon\to0$ we have 
\be
{\cal I}_\nu\ =  {\cal H}[\ \vec u^{(k)}\ ]\ + \ o(1) \ =\ 
 {\cal H}[\ \vec a^{(k)}\ ]
\  +\ {\cal H}[\ \vec b^{(k)}\ ] \ + \ o(1),
\label{eq:boo}\ee
 where we have used that 
  $\vec a^{(k)}$ and $\vec b^{(k)}$ have disjoint supports.
Furthermore,
\be
{\cal H}[\ \vec a^{(k)}\ ]\ \ge\ {\cal I}_{\alpha + o(\varepsilon)}\ {\rm and}\ 
{\cal H}[\ \vec b^{(k)}\ ]\ \ge\  {\cal I}_{\nu -\alpha + o(\varepsilon)},\nn\ee
by definition of ${\cal I}_\theta$
Therefore, taking $k\to\infty$ and $\varepsilon\to0$ we get
\be
{\cal I}_\nu\ \ge\ {\cal I}_\alpha\ +\ {\cal I}_{\nu-\alpha}.
\nn\ee
This contradicts (\ref{eq:subadditivity}).



\begin{thebibliography}{99}


\bibitem{kn:Acevesetal} A.B. Aceves, C. De Angelis, G.G. Luther and A. M.
Rubenchik, {\em Multi-dimensional solitons in fiber arrays},  
Opt. Lett. {\bf 19} (1994) 1186.

\bibitem{kn:ALDRT} A.B. Aceves, G.G. Luther, C. De Angelis, A.M. Rubenchik \& S.K.
Turitsyn, {\em Energy localization in nonlinear fiber arrays: collapse-effect
compressor}, Phys. Rev. Lett. {\bf 75} (1995) 73--76.

\bibitem{kn:Aubry} S. Aubry, {\em Breathers in nonlinear lattices:
Existence, linear stability and quantization}, Physica D {\bf 103}
(1997) 201--250.

\bibitem{kn:BRC} O. Bang, J.J. Rasmussen \& P.L. Christiansen, {\em Subcritical
localization in the discrete nonlinear Schr\"odinger equation with arbitrary
power nonlinearity}, Nonlinearity {\bf 7} (1994) 205--218.

\bibitem{kn:BL} H. Brezis \& E. Lieb, Commun. Math. Phys. {\bf 96}
(1984) 97.

\bibitem{kn:BA}  A.V. Buryak \& N.N. Akhmediev, {\em Stationary pulse propagation in
n-core nonlinear fiber arrays}, IEEE J. Quant. Electr.
{\bf 31} (1995) 682.

\bibitem{kn:CL} T. Cazenave \& P.-L. Lions, {\em Orbital stability of standing
waves for some nonlinear Schr\"odinger equations}, Commun. Math. Phys. {\bf 85}
   (1982) 549--561

\bibitem{kn:colinweinstein} 
 T. Colin \& M.I. Weinstein, {\em On the ground states of
vector nonlinear Schr\"odinger equations}, Ann. Inst. Henri Poincar\'e,
{\bf 65} (1996) 57--79.

\bibitem{kn:Eilbecketal} J.C. Eilbeck, P.S. Lomdahl and A.C. Scott,
{\em The di
screte
self-trapping equation}, Physica D {\bf 16} (1985) 318-338.

\bibitem{kn:experiment} H.S. Eisenberg, Y. Silberberg, R. Morandotti, 
 A.R. Boyd and J.S. Aitchison,  
 {\em Discrete spatial optical solitons in waveguide 
 arrays}, Phys. Rev. Lett. {\bf 81} (1998) 3383.

\bibitem{kn:Flach1} S. Flach, {\em Tangent bifurcation of band edge plane waves,
 dynamical symmetry breaking and vibrational localization}, Physica D {\bf 91} 
 (1996) 223.

\bibitem{kn:Flach2} S. Flach, {\em Breathers on latttices with long range interaction}, Phys. Rev. E {\bf 58} (1998) R4116. 

\bibitem{kn:FKM}  S. Flach, K. Kladko \& R.S. MacKay, {\em  Energy thresholds
for discrete breathers in one-, two-, and three- dimensional lattices},
 Physical Review Letters (1997)

\bibitem{kn:FW} S. Flach \& C.R. Willis, {\em Discrete breathers}, Phys.
Rep. {\bf 295} (1998) 181--264

\bibitem{kn:Friedman} A. Friedman, Partial Differential Equations, Holt 
 Rinehart \& Winston, New York (1969)

\bibitem{kn:GV} J. Ginibre \& G. Velo, {\em  On a class of nonlinear Schr\"odinger
equations}, J. Func. Anal. {\bf 32} (1979) 1--71.

\bibitem{kn:Glassey} R.T. Glassey, {\em  On the blowing up of solutions to the Cauchy
problem for nonlinear Schr\"odinger equations}, J. Math. Phys. {\bf 18} (1977)
1794--1797.

\bibitem{kn:Kato} T. Kato, {\em On nonlinear Schr\"odinger equations}, Ann. Inst. Henri
Poincar\'e, Phys. Th\'eor. {\bf 46} (1987) 113-129.

\bibitem{kn:LST} E.W. Laedke, K.H. Spatschek \& S.K. Turitsyn, {\em Stability of
discrete solitons and quasicollapse to intrinsically localized modes}, Phys.
Rev. Lett. {\bf 73} (1994) 1055--1059.

\bibitem{kn:LSTM} E.W. Laedke, K.H. Spatschek, S.K. Turitsyn \& V.K. Mezentsev,
{\em Analytic criterion for soliton instability in a nonlinear fiber array}, Phys.
Rev. E {\bf 52} (1995) 5549--5554.

\bibitem{kn:MA} R.S. MacKay \& S. Aubry, {\em Proof of existence of
breathers for time-reversible or Hamiltonian networks of weakly coupled
oscillators}, Nonlinearity {\bf 7} (1994) 1623--1643.

\bibitem{kn:PLL}  P.-L. Lions, {\em The concentration compactnes principle in
the calculus of variations I: The locally compact case}, Ann. Inst. Henri
Poincar\'e, Analyse Nonlin\'eaire, {\bf 1} (1984) 223.


\bibitem{kn:MW}  B. Malomed \& M.I. Weinstein, {\em Soliton dynamics in the discrete
nonlinear Schr\"odinger equation}, Phys. Lett. A {\bf 220} (1996) 91.




 


 \bibitem{kn:Strauss} W.A. Strauss, {\em Dispersion of low-energy waves
 for two conservative equations},  
   Archive Rat. Mech. Anal. {\bf 55} (1974) 86--92.

  \bibitem{kn:VPT} 
   V.N. Vlasov, I.A. Petrishchev \& V.I. Talanov, Isv. Rad. {\bf 14}
  (1971) 1353.

\bibitem{kn:W86} M.I. Weinstein {\em Lyapunov stability of ground
states of nonlinear dispersive evolution equations}, Commun. Pure Appl. Math.
{\bf 39} (1986) 51.

\bibitem{kn:WY} M.I. Weinstein \& B. Yeary, {\em Excitation and dynamics of
soliton pulses in optical fiber arrays}, Phys. Lett. A {\bf 222} (1996) 157--162.

\bibitem{kn:nls83} M.I. Weinstein, {\em Nonlinear Schr\"odinger
equations and sharp interpolation estimates}, Commun. Math. Phy. {\bf
87} (1983) 567

\bibitem{kn:survey} M.I. Weinstein, {\em The nonlinear Schr\"odinger
equation - Singularity formation, stability and dispersion},
Contemporary Mathematics {\bf 99} (1989) AMS

 \bibitem{kn:yearythesis} B. Yeary, {\em Coupled nonlinear Schr\"odinger
 equations}, Thesis, 1997, University of Michigan.

\end{thebibliography}
\end{document}